\documentclass[11pt,a4paper]{article}
\pdfoutput=1

\usepackage{jcappub}
\usepackage[latin1]{inputenc}
\usepackage{graphicx}
\usepackage{epsfig}
\usepackage{subfigure}
\usepackage{color}
\usepackage{comment}
\usepackage{slashed}
\usepackage{soul}

\newcommand{\be}{\begin{equation}} 
\newcommand{\ee}{\end{equation}}
\newcommand{\bea}{\begin{eqnarray}} 
\newcommand{\eea}{\end{eqnarray}}

\def\lsim{\mathrel{\raise.3ex\hbox{$<$\kern-.75em\lower1ex\hbox{$\sim$}}}}
\def\gsim{\mathrel{\raise.3ex\hbox{$>$\kern-.75em\lower1ex\hbox{$\sim$}}}}

\usepackage{array}
\newcolumntype{C}[1]{>{\centering\let\newline\\\arraybackslash\hspace{0pt}}m{#1}}

\newcommand{\yr}{{\rm yr}}
\newcommand{\Mpc}{{\rm Mpc}}
\newcommand{\Gpc}{{\rm Gpc}}

\newcommand{\td}{{\rm d}}
\newcommand{\Msun}{M_\odot}

\begin{document}

\title{Gravitational Waves from Primordial Black Hole Mergers}

\author{Martti Raidal,}
\author{Ville Vaskonen}
\author{and Hardi Veerm\"ae}

\affiliation{NICPB, R\"avala 10, 10143 Tallinn, Estonia}  
                            
\emailAdd{martti.raidal@cern.ch}
\emailAdd{ville.vaskonen@kbfi.ee}
\emailAdd{hardi.veermae@cern.ch}

\abstract{
We study the production of primordial black hole (PBH) binaries and the resulting merger rate, accounting for an extended PBH mass function and the possibility of a clustered spatial distribution. Under the hypothesis that the gravitational wave events observed by LIGO were caused by PBH mergers, we show that it is possible to satisfy all present constraints on the PBH abundance, and find the viable parameter range for the lognormal PBH mass function. The non-observation of a gravitational wave background allows us to derive constraints on the fraction of dark matter in PBHs, which are stronger than any other current constraint in the PBH mass range $0.5-16M_\odot$. We show that the predicted gravitational wave background can be observed by the coming runs of LIGO, and its non-observation would indicate that the observed events are not of primordial origin. As the PBH mergers convert matter into radiation, they may have interesting cosmological implications, for example in the context of relieving the tension between high and low redshift measurements of the Hubble constant. However, we find that these effects are suppressed as, after recombination, no more that $1\%$ of dark matter can be converted into gravitational waves.}

\maketitle

%-------------------------------------------------------------------------------
\section{Introduction}
%-------------------------------------------------------------------------------

The binary black hole merger events discovered by the Advanced Laser Interferometer Gravi\-tatio\-nal-Wave Observatory (LIGO) detectors~\cite{Abbott:2016blz,Abbott:2016nmj,TheLIGOScientific:2016pea,Abbott:2017vtc} imply the existence of relatively heavy, $\mathcal{O}(10\, \Msun)$, coalescing black holes (BHs). The merger rate of BH binaries inferred from these events is $12-213\, \Gpc^{-3}\yr^{-1}$ ~\cite{Abbott:2017vtc}. The origin of the BH binaries remains unclear and, although potential astrophysical explanations for the formation of these BH binaries exist~\cite{TheLIGOScientific:2016htt,Belczynski:2016obo}, it is possible that the observed gravitational wave (GW) events emerged from the coalescence of binaries of primordial black holes (PBH)~\citep{Hawking:1971aa,Carr:1974nx,1975ApJ...201....1C,1975A&A....38....5M,CHAPLINE:1975aa}. For recent reviews on PBH physics see e.g. Refs.~\cite{Frampton:2015xza,Garcia-Bellido:2017fdg}. 

The abundance of PBH dark matter (DM) over a broad mass range $10^{-18} - 10^{4}\, \Msun$  is severely constrained~\cite{Carr:2016drx,Carr:2017jsz}, and even if the PBHs make up all the DM, the present binary formation rate~\cite{1989ApJ...343..725Q,Mouri:2002mc} for a uniform PBH distribution has to be enhanced by a factor of $\mathcal{O}(10^{8})$ to reach the merger rates indicated by LIGO~\cite{Clesse:2016ajp}. However, PBH binaries are expected to form also at the onset of structure formation~\cite{Nakamura:1997sm,Ioka:1998nz}, and a merger rate consistent with LIGO can result even if the PBHs make up only a small fraction of DM~\cite{Sasaki:2016jop,Eroshenko:2016hmn}. The stochastic GW background from these binaries can be used to constrain the PBH abundance~\cite{Wang:2016ana}.

In this paper we consider the GW phenomenology of PBH binaries. Our analysis generalises the previous studies~\cite{Nakamura:1997sm,Ioka:1998nz,Eroshenko:2016hmn,Sasaki:2016jop,Wang:2016ana}, as we consider extended PBH mass functions and estimate the effect of PBH clustering. We study PBH binary formation both in the early and in the late Universe, and show that the former generically gives the dominant contribution to the PBH merger rate. Assuming the lognormal PBH mass function, we perform a fit to the LIGO data to determine the presently favoured mass function parameters. We estimate the resulting stochastic GW background and compare it with the sensitivities of GW interferometers. As a result, we obtain the strongest constraints so far on the PBH abundance with masses distributed around $10\Msun,$ and predict a discovery of the GW background by the future runs of LIGO.

To study this scenario further, we consider the loss of DM energy density due to the PBH coalescence. As the PBH mergers convert part of their mass into GWs, the measurements of late time cosmological evolution impose an upper bound on their merger rate. An intense period of merging after recombination could even relieve the tension~\cite{Enqvist:2015ara,Berezhiani:2015yta,Chudaykin:2016yfk,Poulin:2016nat} between the low and high redshifts measurements of the Hubble constant~\cite{Riess:2016jrr,Hildebrandt:2016iqg,Joudaki:2016kym,Bernal:2016gxb}. However, we show that this possibility is not supported by the merger rates predicted by our analysis.

The paper is organised as follows. In Sec.~\ref{sec:formation} we review the PBH binary formation mechanisms with an extended PBH mass function. In Sec.~\ref{sec:LIGO} we discuss the LIGO observations, show that the PBH mergers can account for the LIGO signal, and perform a fit of the mass function to the observed LIGO events. In Sec.~\ref{sec:GWs} we calculate the stochastic GW background from PBH mergers, and derive the constraints from non-observation of the background by LIGO. In that section we also study the change in DM energy density due to PBH mergers. We summarise our key conclusions in Sec.~\ref{sec:conclusions}.

%-------------------------------------------------------------------------------
\section{Review of PBH binary formation}
\label{sec:formation}
%-------------------------------------------------------------------------------

There are two primary mechanisms for PBH binary formation. They can be classified by the cosmological epoch in which they occur. First, early formation of gravitationally bound objects is expected to coincide roughly with matter-radiation equality. At that time the PBH peculiar velocities are negligible and nearby PBHs will begin to fall towards each other. To describe this process we adopt the three-body approximation~\cite{Nakamura:1997sm,Ioka:1998nz,Sasaki:2016jop}, according to which the two closest PBHs form a binary while the tidal forces from the third PBH prevent a head-on collision.  Ref.~\cite{Eroshenko:2016hmn} extended the analysis of Ref.~\cite{Sasaki:2016jop} by considering also tidal forces from inflationary perturbations. For simplicity we will neglect this contribution, but note that it can slightly increase the PBH merger rate and consequently tighten the constraints shown in Sec.~\ref{sec:GWs}.

Second, in the late Universe well within the non-linear phase of structure formation, PBH binaries can be created by close encounters of PBHs. A bound system is formed if they lose enough kinetic energy either through friction with the environment or by the emission of GWs~\cite{1989ApJ...343..725Q,Mouri:2002mc}. In the context of LIGO, GWs from the second process have been studied in Refs.~\cite{Bird:2016dcv,Clesse:2016ajp,Kovetz:2017rvv}. We remark, that GWs may also be generated in the late Universe by close BH encounters even if the BHs do not form a binary~\cite{Garcia-Bellido:2017qal}.

Our analysis differs from the previous ones as we consider binary formation for a population of PBHs with an extended mass function $\psi(m)$. We normalise the mass function to the fraction of PBH DM, $f_{\rm PBH} \equiv \Omega_{\rm PBH}/\Omega_{\rm DM}$, that is
\be
	f_{\rm PBH} = \int_0^\infty \psi(m) \td m \,.
\ee
One of the simplest choices for $\psi(m)$ is the lognormal distribution of the form
\be\label{def:psi}
	\psi(m) = \frac{f_{\rm PBH}}{\sqrt{2\pi}\sigma m} \exp\left(-\frac{\log^2(m/m_c)}{2\sigma^2}\right) \,,
\ee
where $m_c$ denotes the peak mass of $m\psi(m)$, and $\sigma$ is the width of the mass spectrum. The limiting case $\sigma \to 0$ is the monochromatic mass function centered around $m_{c}$, $\psi(m) = f_{\rm PBH}\delta(m-m_{c})\,$. An approximately lognormal mass function arises naturally in various models of PBH production~\cite{Dolgov:1992pu,Clesse:2015wea,Bartolo:2016ami,Blinnikov:2016bxu,Green:2016xgy,Kannike:2017bxn}. In case of inflationary production of PBHs, if the slow-roll approximation is violated during some period of inflation, as may happen in the case of single field double inflation~\cite{Germani:2017bcs,Kannike:2017bxn,Motohashi:2017kbs,Garcia-Bellido:2017mdw,Ezquiaga:2017fvi,Bezrukov:2017dyv}, the resulting PHB mass function generically deviates from the lognormal form. The precise form of the mass function must then be derived case-by-case. For example, it was shown in Ref.~\cite{Kannike:2017bxn} that wider mass functions usually develop an extended low mass tail when compared to the lognormal form. Another common choice is the power law mass function~\cite{Carr:1975qj,Carr:2017edp,Carr:2017jsz,Magee:2017vkk} which we do not consider in this paper.

%-------------------------------------------------------------------------------
\subsection{Binary formation in the early Universe}

Binary formation starts when the Newtonian gravitational attraction of nearby PBHs overtakes the effect of the cosmic expansion. To estimate when this happens, consider a pair of nearby PBHs with masses $m_{1}$ and $m_{2}$, and a physical spatial separation $r$, which is assumed to be much smaller than the size of the Hubble horizon. The peculiar velocities in the radiation dominated era are damped by Hubble friction, so the PBH initial velocities can be ignored. In the Newtonian approximation, the physical distance of a pair of point masses in the Friedmann-Robertson-Walker background obeys the equation of motion  $\ddot r = - G M r^{-2} + (\ddot a/a) r$, where $M \equiv m_{1} +m _{2}$ denotes the total mass of the pair and $a$ is the scale factor.  The Friedmann equations imply that $\ddot a/a  = - (1+3w)H^2/2,$ where $w$ is the equation of state parameter, and $H$ is the Hubble parameter. Both $w$ and $H$ depend on the background energy density $\rho_{\rm bg},$ which is obtained by subtracting the energy density in PBHs from the total local energy density. 

When the motion of the PBH pair is dominated by the expansion, their physical distance scales roughly as $r \propto a$. The moment when the pair decouples from the cosmic expansion is estimated by requiring that the Newtonian force dominates in the equation of motion. This corresponds to $G M > (1+3w) H^2 r^3/2$. In case the background is dominated by matter, $H^2r^3$ is constant. It follows that decoupling can occur only in a radiation dominated background, where the decoupling condition reads
\be\label{eq:cond_dc}
	\frac12 M > \frac{4\pi}{3} \rho_{\rm bg} r^{3} \,.
\ee
This condition has a simple interpretation: decoupling from the cosmic expansion takes place when the average mass of the PBHs forming the binary is larger than the background mass enclosed in a sphere of radius $r$. This interpretation of the condition \eqref{eq:cond_dc} was used in Ref.~\cite{Nakamura:1997sm}. 

The background energy density can be approximated as $\rho_{\rm bg} \approx (a_{\rm eq}/a)^4 \rho_{\rm eq}/2$, where $\rho_{\rm eq}$ is the energy density of the Universe at matter-radiation equality $a_{\rm eq}$.\footnote{The scale factor is normalised such that today $a=1$.} Then solving $a$ from \eqref{eq:cond_dc} gives the moment of decoupling,
\be
	a_{\rm dc} \approx a_{\rm eq} \left(\frac{x}{\tilde{x}}\right)^{3} \,,
\ee
where $x \equiv r/a$ is the comoving distance, and we defined the scale
\be
	\tilde{x}^3 \equiv \frac{3}{4\pi} \frac{M}{a_{\rm eq}^3\rho_{\rm eq}} \,.
\ee

Depending on the fraction of DM in PBHs, two qualitatively different cases exist. First, in case most of the DM consists of PBH, the distance $\tilde{x}$ corresponds roughly to the maximal distance of PBHs, as larger separations are exponentially unlikely. Second, if PBHs constitute only a subdominant fraction of DM, the background is matter dominated after matter-radiation equality. So, by the above arguments, decoupling should take place before that, i.e. $a_{\rm dc}<a_{\rm eq}$. In both cases $x < \tilde{x}$. A slightly different argument led to a similar condition in Ref.~\cite{Sasaki:2016jop}.

To estimate the parameters of the forming PBH binary we follow Refs.~\cite{Nakamura:1997sm, Ioka:1998nz}. In the three-body approximation two PBHs fall towards each other while the tidal forces from the PBH closest to the centre of mass of the pair provides angular momentum that may prevent their head-on collision. Let us denote the mass of the third PBH by $m_3$ and its comoving separation from the PBH pair by $y>x$. For this system the semi-major axis $r_a$ and the semi-minor axis $r_b$ of the binary are estimated by
\be\label{eq:ra,rb}
	r_{a} = \alpha\, x a_{\rm dc} \,, \qquad
	r_{b} = \beta\,  \frac{2m_{3}}{M}  \left(\frac{x}{y} \right)^{3} r_{a} \,,
\ee
where $\alpha$ and $\beta$ are $\mathcal{O}(1)$ numerical constants. Throughout this paper $\alpha = \beta = 1$ is used, which was shown to be an adequate approximation in Ref.~\cite{Sasaki:2016jop}. The coalescence time for such a binary is~\cite{Peters:1964zz}
\be \label{eq:BHtime}
	\tau \approx \frac{3}{85} \frac{r_{a}^{-3} r_{b}^{7}}{\eta\,(G M)^{3}} = \tilde{\tau}  \left(\frac{x}{\tilde{x}}\right)^{37} \left(\frac{y}{\tilde{x}}\right)^{-21},
\ee
where  $\eta \equiv m_{1}m_{2}/M^{2}$ and
\be
	\tilde{\tau} \equiv \frac{384}{85} \, \frac{\alpha^4 \beta^7 a_{\rm eq}^4  m_3^7 \tilde{x}^4}{ G^3 \eta M^{10}} \,,
\ee
gives the maximal coalescence time because $x < \tilde{x},y$. Note, that $\tilde{\tau}$ exceeds the age of the Universe by several orders of magnitude.  Eq.~\eqref{eq:BHtime} corresponds to the maximal eccentricity limit, with $r_{a} \gg r_{b}$, but remains a good approximation for low eccentricities also. 

Having determined the properties of the binary that will be created from the initial configuration of PBHs with masses $m_{1}$, $m_{2}$, $m_{3}$ and comoving distances $x$, $y$, the distribution of the binary parameters, e.g. the coalescence time, follow from the mass function $\psi(m)$ and the spatial distribution of the PBHs. Assuming an isotropic distribution, the average number of PBHs in a mass interval $(m,m+\td m)$ and within a spherical shell $(x,x+\td x)$ surrounding a PBH is
\be \label{eq:Nxm}
	\td N(x,m) = 4\pi x^2 \, \rho_{\rm DM} \frac{\psi(m)}{m} \, (1 + \xi(x))\td x \td m \,,
\ee
where $\rho_{\rm DM}$ is the present dark matter energy density and $\xi(x)$ is the PBH two point function.  Notice, that the fraction of DM in PBHs is included in the definition of the mass function $\psi$.

In the absence of non-gaussianity, the PBH spatial distribution will follow the overall DM density perturbations~\cite{Tada:2015noa}. This is easily understood in a time slicing for which the superhorizon density perturbations vanish. In that case, if a density fluctuation with wavenumber $k$ enters the horizon, the probability that it collapses to a black hole is clearly the same in each Hubble patch~\cite{Young:2014ana}, but the collapses take place at slightly different times. Let $\delta t$ describe this time delay. Then, in the linear regime $\delta \rho_{\rm PBH} \propto \delta t$, so the energy density will follow the overall density perturbations in the spatially flat slicing. 

The primordial curvature perturbations must be less than unity in order to avoid overproduction of PBHs. As these perturbations grow very slowly in a radiation dominated Universe, $\xi(x)$ is not expected to exceed unity at the time of matter-radiation equality. Nevertheless, small scale density perturbations may grow significantly, if there is an extended matter dominated period between PBH formation and big bang nucleosynthesis~\cite{Dolgov:2011cq}. Also, small scale non-gaussianity can introduce  isocurvature perturbations and thereby produce fluctuations in the PBH number density which are larger than the background perturbations~\cite{Tada:2015noa}, or the presence of closed domain walls may lead to formation of PBH clusters~\cite{Rubin:2001yw,Khlopov:2002yi,Khlopov:2004sc}. So it is possible that $\xi(x)$ may be large enough to significantly affect binary production. 

To obtain a rough quantitative estimate of the effect from PBH clustering, we assume  an idealised two point function that is constant at comoving distances smaller than $\tilde x$, 
\be\label{eq:deltadc}
1 + \xi(x) \approx \delta_{\rm dc}\,, \quad {\rm if }\,\, x < \tilde{x} \,.
\ee
The quantity $\delta_{\rm dc}$ is the local density contrast at $a_{\rm dc}$ at scales smaller than $\tilde{x}$. Strong clustering thus corresponds to large values of the density contrast, $\delta_{\rm dc}\gg1$. Using a more realistic two-parameter description, e.g. a power law, for the two point function will slightly modify the lifetime distribution of the binaries. However, we have checked numerically that this effect is expected to modify our results by less than an order of magnitude.

The comoving number density of binaries resulting from the three PBH configurations described above is
\be \label{def:n3}
	\td n_{3}(x,y) = \frac{1}{2} \left( \rho_{\rm DM} \frac{\psi(m_{1})}{m_{1}} \, \td m_{1} \right) \, \left( e^{-N(y)}  \, \td N(x,m_{2}) \, \td N(y,m_{3}) \right) \,,
\ee
where the first bracket denotes the density of PBHs in the mass range $(m_{1},m_{1}+\td m_{1})$, and the second bracket gives the probability that they belong to the three body configuration with the specified parameters. The factor $1/2$ avoids double counting of binaries, and $N(y) \equiv \int  \td N(y,m)$ is the expected number of PBHs surrounding a PBH in the sphere of comoving radius $y$. The exponential term arises from Poisson statistics of the PBH number and corresponds to the requirement that there are no other PBHs within the radius $y$. We remark that in case the mass function is monochromatic, the energy density of binaries is given by $\rho_{\rm PBH} (1 - e^{-N(\tilde{x}) })$. For $\delta_{\rm dc} f_{\rm PBH} \ll 1$ the energy density of PBH binaries is proportional to $\delta_{\rm dc } f_{\rm PBH}^2$.

The mass dependent differential merger rate per comoving volume can be expressed as
\be \label{rateR30}
	\td R_3(t) = \int_0^{\tilde{x}} \td x \int_x^\infty \td y \, \frac{\partial^2 n_3(x,y)}{\partial x \partial y} \, \delta(t-\tau(x,y,m_{i})).
\ee 
We do not assume an upper bound for $y$, so the integration includes regions where the approximation \eqref{eq:deltadc} does not hold. The resulting error is expected to be small  due to the exponential suppression factor in Eq.~\eqref{def:n3}. Plugging the coalescence time~\eqref{eq:BHtime} into Eq.~\eqref{rateR30} gives
\be \label{rateR3}
\begin{aligned}
	\td R_3(t) 
	=& \frac{9}{296\pi} \, \frac{1}{\tilde{\tau}} \left(\frac{t}{\tilde{\tau}}\right)^{-\frac{34}{37}}\left( \Gamma\left[\frac{58}{37}, \tilde{N} \left(\frac{t}{\tilde{\tau}}\right)^\frac{3}{16}\right] - \Gamma\left[ \frac{58}{37},\tilde{N} \left( \frac{t}{\tilde{\tau}} \right)^{-\frac{1}{7}} \right] \right) 
	\\ & \times \tilde{x}^{-3} \delta_{\rm dc}^{-1}\tilde{N}^{53/37} \bar{m}^3 \frac{\psi(m_{1})}{m_{1}} \frac{\psi(m_{2})}{m_{2}}  \frac{\psi(m_{3})}{m_{3}} \, \td m_1 \td m_2 \td m_3\,,
\end{aligned}
\ee 
where we defined
\be
	\bar{m} \equiv \left(\int \td m \frac{\psi(m)}{m}\right)^{-1} \,,
\ee
and $\tilde{N} \equiv N(\tilde{x})= \delta_{\rm dc}\Omega_{\rm DM,eq} M/\bar{m}$, with the DM density parameter at matter-radiation equality given by $\Omega_{\rm DM,eq}=0.42$. The average PBH mass is given by $\bar{m}f_{\rm PBH}$. 

The rate~\eqref{rateR3}  vanishes at $t=\tilde{\tau}$, as expected, since $\tilde{\tau}$ gives the maximal coalescence time. The effect of PBH spatial clustering does not influence $\tilde{\tau}$ and is mostly encoded in $\tilde{N}$. The larger $\tilde{N}$ is, the more likely are shorter coalescence times. However, $\tilde{\tau}$ can be modified by fluctuations in the background energy density, which we have neglected.

The rate~\eqref{rateR3} gets corrections from various dynamical effects which we neglect in our analysis. Effects affecting binary formation include the dependence on the angular coordinates of the third PBH, the possibility of a $3$-body collision in case the third PBH is bound to the pair, $n$-body effects arising from other PBHs surrounding the pair, and the effect of the initial velocity of the BHs. These effects will generally introduce a $\mathcal{O}(1)$ correction to the rate~\eqref{rateR3} and may be accounted for by modifying the values of the fudge factors $\alpha$ and $\beta$ introduced in Eq.~\eqref{eq:ra,rb}~\cite{Ioka:1998nz}.

The evolution of the binaries can also have an influence on properties of the binary population and thereby the rate \eqref{rateR3}. A binary can be disrupted or broken apart by the gravitational pull from a passing PBH. This mechanism is similar to the survival of wide star binaries~\cite{Monroy-Rodriguez:2014ula} so binaries with larger separations or comprising lower mass PBHs will be affected more by it. In case the PBH have subsolar masses, also stars may perturb the binaries. The probability of disruption was estimated to be $\mathcal{O}(10^{-7})$ in a galactic halo~\cite{Sasaki:2016jop}. However, it has been argued in Ref.~\cite{Kovetz:2017rvv} that this result may underestimate the overall disruption rate. Disruption may also be enhanced in regions with a higher PBH density contrast.

%-------------------------------------------------------------------------------
\subsection{Binary formation in the late Universe}

In the late Universe the PBHs are expected to have large peculiar velocities which renders the binary formation unlikely. Nevertheless, a PBH binary can be formed by two-body interactions via energy loss due to gravitational radiation, given the two PBHs pass each other at a sufficiently close distance. In the Newtonian approximation, the velocity averaged cross section for the PBH binary formation via this mechanism is~\cite{1989ApJ...343..725Q,Mouri:2002mc}
\be \label{eq:vsigma}
	\langle v_{\rm rel}\sigma_{\rm mer} \rangle = A \, G^2 M^2 \left(\frac{\eta}{v_{\rm PBH}^9} \right)^{2/7} \,,
\ee
where $v_{\rm rel}$ denotes the relative velocity of the PBHs, $v_{\rm PBH}$ is a characteristic velocity of the PBHs, and $A$ is a numerical constant which depends on the shape of the velocity distribution, given it has the functional from $f(v/v_{\rm PBH}$). We use $A \approx 12.9$, which corresponds to the Maxwellian velocity distribution $P(v) \propto v^{2} \exp\left(-v^{2}/v_{\rm PBH}^{2}\right)$. In this case, $v_{\rm PBH}$ is related to the PBH velocity dispersion, which is expected to be of order $10 - 100\, {\rm km}/{\rm s}$. 

The differential merger rate per comoving volume corresponding to \eqref{eq:vsigma} reads
\be\label{rateR2}
	\td R_2(t) = \delta(t) \rho_{\rm DM}^2 \, \langle v_{\rm rel}\sigma_{\rm mer} \rangle \, \frac{\psi(m_1)}{m_1} \frac{\psi(m_2)}{m_2} \, \td m_1 \td m_2 \,,
\ee 
where, as in the previous section, we included the DM density contrast $\delta$ to quantify the effect of clustering~\cite{Clesse:2016vqa,Clesse:2016ajp}. Since density perturbations grow significantly after the matter-radiation equality, it is expected that the density contrast today, $\delta_0$, is much larger than $\delta_{\rm dc}$. 

PBHs that are clumped into smaller systems tend to have a smaller velocity dispersion. So clustering of PBHs will generally enhance the rate \eqref{rateR2} not just through an increase of $\delta_{0}$ but also by decreasing $v_{\rm PBH}$ in the velocity averaged cross section \eqref{eq:vsigma}.  In the rate~\eqref{rateR2} we also neglected the possible spatial dependence of the PBH mass distribution caused by mass segregation, which tends to expel the low mass PBHs from regions with a high density contrast.

%-------------------------------------------------------------------------------
\section{Implications from LIGO observations}
\label{sec:LIGO}
%------------------------------------------------------------------------------- 

In this section we study the possibility that the GW events observed by LIGO have a purely primordial origin. We compute the PBH merger rates from both the mechanisms described in the previous section and perform a fit to the parameters of the lognormal mass function~\eqref{def:psi}.

%-------------------------------------------------------------------------------
\subsection{The present merger rate}

To analytically estimate the merger rate of PBHs formed by the two different mechanisms we will first assume a monochromatic mass function. The present rate of PBH mergers from binaries formed in the early Universe can be approximated by
\be \label{R3approx}
	R_3(t_0) \approx 5.1\times10^4 \delta_{\rm dc}^{16/37} f_{\rm PBH}^{53/37} \left(\frac{m_c}{30\Msun}\right)^{-32/37} \Gpc^{-3} \yr^{-1}\,,
\ee
given that $0.007 (m_c/30\Msun)^{5/21}\lsim \delta_{\rm dc} f_{\rm PBH}\lsim 9000 (m_c/30\Msun)^{5/16}$. In case these conditions are violated, this approximation will overestimate the rate. The merger rate of PBH binaries formed in the late Universe reads
\be
	R_2(t_0) \approx 3.7\times10^{-7} \delta_0 f_{\rm PBH}^2 \left(\frac{v_{\rm PBH}}{10\,{\rm km/s}}\right)^{-11/7} \Gpc^{-3} \yr^{-1}\,.
\ee
For example, if  $f_{\rm PBH}=1$, $\delta_{\rm dc} = 1$, $m_c=30\Msun$ and $v_{\rm PBH}=10\,{\rm km/s}$, the present day DM density contrast has to be $\delta_0 \gtrsim 10^{11}$ for $R_2$ to exceed $R_3$. This bound will become even stricter if the PBHs do not make up all of the DM, because $R_2$ is more sensitive to $f_{\rm PBH}$ than $R_3$. So $R_3$ will generally dominate over $R_2$. An exception to this rule occurs if the early density contrast $\delta_{\rm dc}$ is extremely high. In that case, most of the binaries have merged within a timescale much shorter than the lifetime of the Universe, and $R_2$ would thus be the only remaining and dominant source of PBH mergers.

\begin{table}
\begin{center}
\caption{The masses of the BH binary constituents $m_{1}$ and $m_{2}$ for the GW events observed by LIGO together with the symmetric mass ratios $\eta$ of the binaries.}
\label{tab:LIGO}
\begin{tabular}{lC{1.6cm}C{1.6cm}c}
   & $m_1/\Msun$ & $m_2/\Msun$ & $\eta$ \\
  \hline
  GW150914~\cite{Abbott:2016blz} & $36\substack{+5 \\ -4}$  & $29\substack{+4 \\ -4}$ & $0.247\substack{+0.005 \\ -0.005}$ \\
  GW151226~\cite{Abbott:2016nmj} & $14.2\substack{+8.3 \\ -3.7}$  & $7.5\substack{+2.3 \\ -2.3}$ & $0.226\substack{+0.046 \\ -0.028}$ \\
  LVT151012~\cite{TheLIGOScientific:2016pea} & $23\substack{+18 \\ -6}$  & $13\substack{+4 \\ -5}$ & $0.231\substack{+0.054 \\ -0.030}$ \\
  GW170104~\cite{Abbott:2017vtc} & $31.2\substack{+8.4 \\ -6.0}$  & $19.4\substack{+5.3 \\ -5.9}$ & $0.236\substack{+0.021 \\ -0.020}$ 
\end{tabular}
\end{center}
\end{table}

For the rate $R_3(t_0)$ to be consistent with the rate indicated by the LIGO observations, $12-213\, \Gpc^{-3}\yr^{-1}$ ~\cite{Abbott:2017vtc}, the PBH DM fraction has to be at the percent level. For example, a monochromatic mass function with $m_c=30\Msun$ implies $f_{\rm PBH} = 0.0033-0.022$ for $\delta_{\rm dc}=1$. This PBH fraction is  slightly larger than the one obtained in Ref.~\cite{Sasaki:2016jop}. In Ref.~\cite{Sasaki:2016jop} an upper bound $y<\tilde{x}$ is required, whereas we integrate up to arbitrarily large $y$ and the exponential factor in Eq.~\eqref{def:n3}, which is absent in Ref.~\cite{Sasaki:2016jop}, suppresses the integral at large $y$.

A larger early density contrast, $\delta_{\rm dc} > 1$, requires  the fraction $f_{\rm PBH}$ to be even smaller. Fixing the rate and the PBH mass in the approximation \eqref{R3approx} we find that $f_{\rm PBH} \propto \delta_{\rm dc}^{-16/53}$, so the fraction of DM in PBHs scales relatively weakly with the density contrast. 

We remark that in the above scenario only a small fraction of PBHs is in binaries. For example, if $f_{\rm PBH} = 0.0033-0.022$, then only the fraction $0.0027-0.018$ of PBHs is in binaries. For larger $f_{\rm PBH}$ this fraction increases and, according to \eqref{def:n3}, it is close to unity for $f_{\rm PBH} =1$.

%-------------------------------------------------------------------------------
\subsection{The extended mass function}

Provided that all BHs observed by LIGO are primordial, the masses of the merging BHs observed so far indicate an extended PBH mass function. We perform a maximum likelihood fit to the masses listed in Table~\ref{tab:LIGO} assuming a lognormal mass function. Accounting for the experimental uncertainties and the finite sensitivity range of LIGO, the log-likelihood function reads
\be
\ell = \sum_j \ln \int_{m_{\rm min}}^{m_{\rm max}} \td m\, p_j(m_j | m) \frac{ p_\psi(m)}{P_\psi(m_{\rm min}<m<m_{\rm max})} \,,
\ee 
where $p_\psi(m)\propto \psi(m)/m$ is the probability that the PBH has mass $m$, and $p_j(m_j | m)$ is the probability that we observe a PBH mass $m_j$ given that the PBH has mass $m$. We take it to be a Gaussian distribution with mean $m$ and variance $\sigma_j^2$. The masses $m_j$ and the standard deviations $\sigma_j$ correspond to the masses and their maximal errors given in Table~\ref{tab:LIGO}. To roughly estimate the effect of the LIGO sensitivity, we take $(m_{\rm min}=7\Msun, m_{\rm max}=50\Msun)$ as the range of PBH masses detectable by LIGO. This is accounted by the normalisation factor $P_\psi(m_{\rm min}<m<m_{\rm max})$ giving the probability of finding a PBH in the mass range $(m_{\rm min},m_{\rm max})$. The $p_\psi(m)/P_\psi$ term thus corresponds to the conditional probability of finding a PBH of mass $m$ given that it lies in the mass range $(m_{\rm min},m_{\rm max})$. We emphasise that this is a rough estimate. In a more careful analysis one should compare the GW signal from the PBH merger to the LIGO sensitivity curve given, e.g., in Ref.~\cite{TheLIGOScientific:2016wyq}.

\begin{figure}
\begin{center}
\includegraphics[height=.42\textwidth]{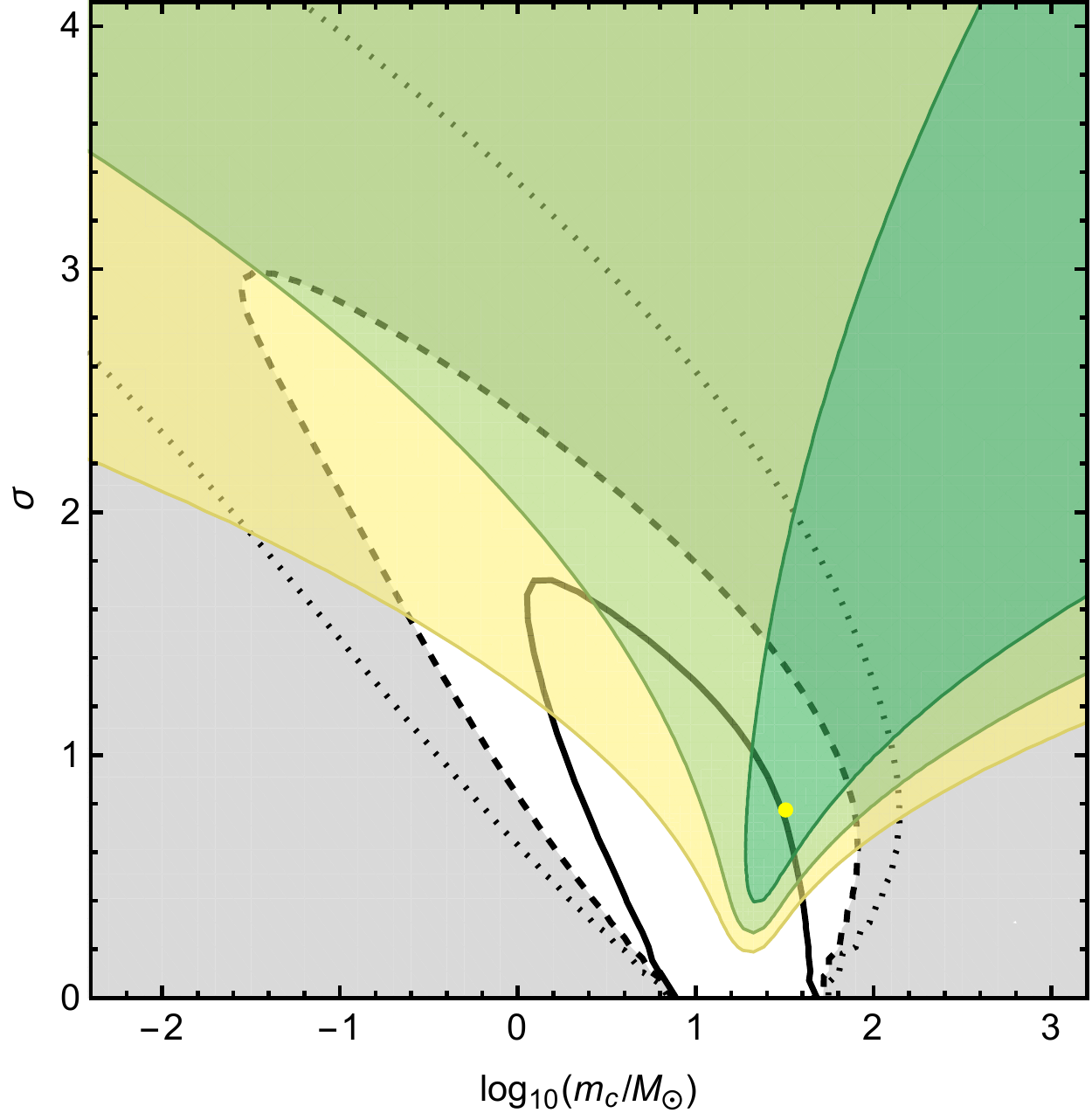} \hspace{4mm}
\includegraphics[height=.44\textwidth]{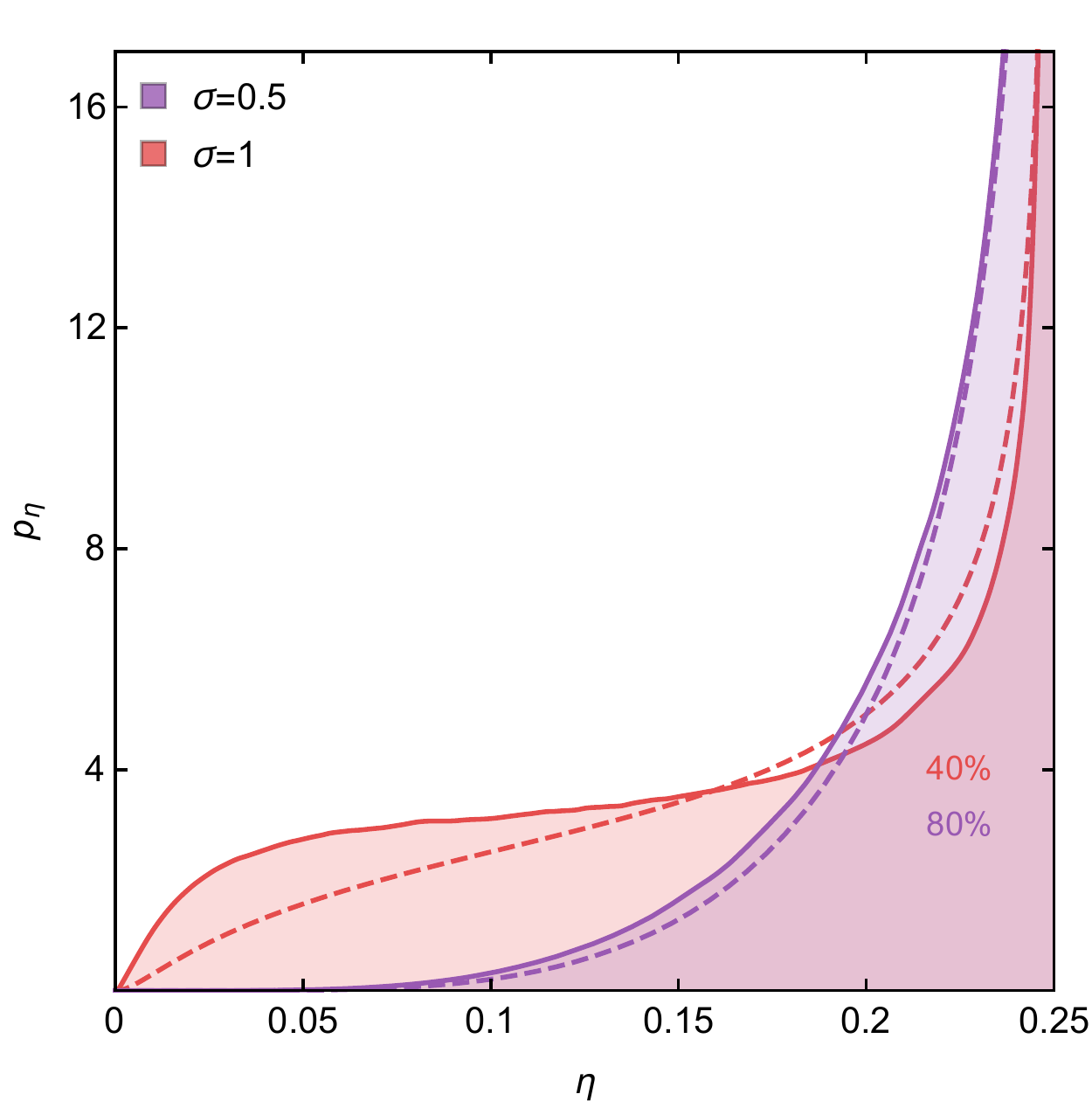}
\end{center}
\caption{\emph{Left panel:} The results of maximum likelihood fit to the LIGO GW events for the lognormal mass function parameters $m_c$ and $\sigma$.
The dot depicts the best fit value and the green, light green and yellow contours show, respectively, the relative likelihoods $\ell-\ell_{\rm max} = -1/2, -4/2, -9/2$. The contours of maximum allowed fraction of DM in PBHs in the LIGO sensitivity range $f_{\rm max}(m_{\rm min}<m<m_{\rm max})=10^{-2},10^{-3},10^{-4}$ are shown by the black solid, dashed and dotted lines, respectively. \emph{Right panel:} The probability density function of the symmetric mass ratio of the binary for lognormal PBH mass function. The solid lines correspond to the binaries merging today, and the dashed lines to all binaries formed. The numbers denote the fraction of binaries with $\eta>0.2$ for the solid lines.}
\label{LIGOplot}
\end{figure}

The best fit  result, $(m_c,\sigma) = (33\Msun,0.8),$ and the relative likelihood contours for $\ell-\ell_{\rm max} = -1/2, -4/2, -9/2$ are shown in the left panel of Fig.~\ref{LIGOplot}. As a representative example, consider a lognormal mass function with $m_c  = 30\Msun$ and $\sigma = 1$. The LIGO rate in the mass range $(m_{\rm min},m_{\rm max})$ for this mass function is recovered if $f_{\rm PBH} = 0.0045 - 0.024$. 

The black lines in the left panel of Fig.~\ref{LIGOplot} depict the maximal allowed fraction of DM in PBHs in the mass range $(m_{\rm min},m_{\rm max})$. These are obtained by first calculating the maximal allowed PBH abundance for given $m_c$ and $\sigma$ with the method described in Ref.~\cite{Carr:2017jsz} and then integrating the mass function to obtain the PBH mass fraction in the the LIGO sensitivity range. We consider all constraints used in Ref.~\cite{Carr:2017jsz}\footnote{We included the dynamical constraints from Ref.~\cite{Carr:2017jsz} and, as in Ref.~\cite{Carr:2017jsz}, we omitted the accretion constraints~\cite{Carr:1979,Gaggero:2016dpq,Inoue:2017csr} due to their strong dependence on largely uncertain BH accretion physics.}. 
 
As the LIGO sensitivity range is relatively narrow, wider mass functions will be less constrained, as can be seen from Fig.~\ref{LIGOplot}. However, if $m_{c}$ lies far from the mass range $(m_{\rm min},m_{\rm max})$, the fraction of PBH in this range, and therefore also the merger rate, will decrease. Consistence with the LIGO rate excludes the gray region in Fig.~\ref{LIGOplot}, because it requires that the fraction of DM in PBHs in that mass range is larger than $0.1\%$. The values of $m_c$ and $\sigma$ compatible with the LIGO observations are therefore within the region enclosed by the light green and the dashed black curves. Viable lognormal mass functions have a width   in the range $\sigma \approx 0.3 - 3$ while $m_{c} \approx 0.1 - 60 \Msun$.

Given the mechanism of binary formation, it is possible to compare the overall PBH mass function with the mass distribution of the PBHs in binaries merging today, which was, in fact, estimated above. We will consider the distribution of the symmetric mass fraction $\eta\equiv m_{1}m_{2}/(m_1+m_2)^2$, that ranges from $0$, for maximally asymmetric masses, to $1/4$, signifying that  both masses are equal. If the binaries are combined randomly, the $\eta$ distribution is
\be
	p_{\eta,{\rm dc}} \propto \int \td m_{1} \td m_{2} \frac{\psi(m_{1})}{m_{1}} \frac{\psi(m_{2})}{m_{2}} \delta\left(\eta - \frac{m_{1}m_{2}}{(m_1+m_2)^2}\right).
\ee
Assuming the three body mechanism, this corresponds to the $\eta$ distribution of the PBHs in the early Universe when they were formed. However, in general this does not provide the correct description of the binaries that are merging today. Instead, the $\eta$ distribution for these binaries reads
\be
	p_{\eta,0} \propto \int \td R_3(t_{0}) \, \delta\left(\eta - \frac{m_{1}m_{2}}{(m_1+m_2)^2}\right).
\ee
In this estimate we assume that PBH binaries evolve uninterrupted from their formation until they merge and thus neglect the disruption of binaries by surrounding compact objects. 

Both $p_{\eta,{\rm dc}}$ and $p_{\eta,0}$ are shown in the right panel of Fig.~\ref{LIGOplot}. The $\eta$ distribution of randomly composed PBH pairs and the binaries whose mergers we may observe today are depicted by the dashed and the solid lines, respectively. Compared to random pairing, $\eta$ today is seen to be less peaked at $\eta = 1/4$, so equal mass binaries are less likely. This effect gets weaker the narrower the mass function is, e.g. for $\sigma = 0.5$ it is practically negligible. If this is not the case, the mass distribution of merging PBHs will generally not coincide with the overall mass function.

The values of $\eta$ for the LIGO observations are given in Table~\ref{tab:LIGO}. All of them are close to the maximal value $\eta=1/4$. This may be an effect of experimental sensitivity, as the detection of mergers with larger $\eta$ is more likely because they produce a stronger signal.

%-------------------------------------------------------------------------------
\section{Gravitational waves from PBH mergers}
\label{sec:GWs}
%-------------------------------------------------------------------------------

In this section we estimate the stochastic GW background from PBH mergers and the DM energy density loss into GWs. The merger rate and the constraints on the GW background from binary mergers can constrain the PBH abundance~\cite{Wang:2016ana,Kovetz:2017rvv}. Also, the cosmological evolution of the late time Universe constrains the change of DM energy density, which in turn results in a bound on the PBH merger rate. We note that the stochastic GW background from PBH binary mergers for an extended mass function has been studied in Ref.~\cite{Bartolo:2016ami}. However, they did not consider binary formation but used a fixed rate.

%-------------------------------------------------------------------------------
\subsection{Constraints from the non-observation of the GW background}

\begin{figure}
\begin{center}
\includegraphics[height=.44\textwidth]{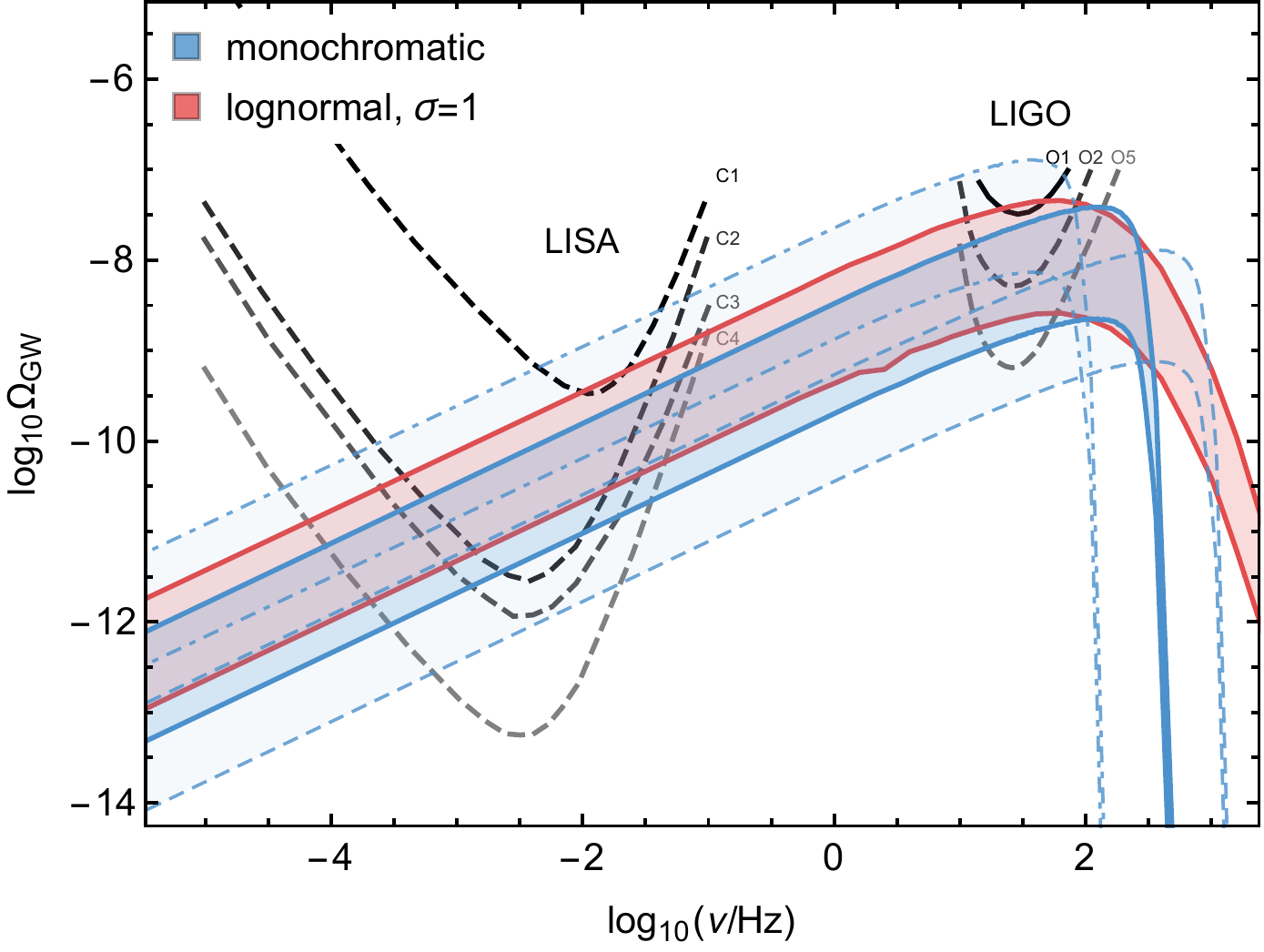}
\end{center}
\caption{The predicted stochastic GW background from the PBH mergers for the monochromatic (blue) and lognormal (red) mass functions. The fraction of DM in PBHs is chosen such that the merger rate in the LIGO sensitivity range today is $12\, \Gpc^{-3} \yr^{-1}$ for the lower lines and $213\, \Gpc^{-3} \yr^{-1}$ for the upper lines, corresponding to $f_{\rm PBH}\sim0.001-0.01$. For the solid lines $m_c=30\Msun$, and for the dashed and dot-dashed lines $m_c=10,100\Msun$, respectively. The black solid line shows the sensitivity of the first LIGO observing run (O1), and the gray dashed lines (O2, O5) below it  show the expected sensitivities of the next observing runs~\cite{TheLIGOScientific:2016wyq}. The dashed grey lines on the left (C1-C4) show the sensitivities of different configurations of LISA~\cite{Caprini:2015zlo}.}
\label{GWspectrum}
\end{figure}

PBH mergers occurring with a differential rate $\td R(z)$ yield a stochastic GW background with the spectrum~\cite{Phinney:2001di,TheLIGOScientific:2016wyq}
\be\label{eq:OGW}
	\Omega_{\rm GW}(\nu) = \frac{\nu}{ \rho_c} \int \td z \, \td R(z) \frac{1}{(1+z) H(z)}  \frac{\td E_{\rm GW}(\nu_r)}{\td \nu} \,,
\ee
where $\rho_c=3H_0^2/8\pi G$ denotes the critical density, $\nu_r=(1+z)\nu$ is the redshifted source frequency and $H(z) = H_0\sqrt{\Omega_\gamma (1+z)^4 + \Omega_m (1+z)^3 + \Omega_\Lambda}$ is the $\Lambda$CDM Hubble parameter at redshift $z$. We neglect the effect of production of GWs on evolution of $H(z)$ and use $\Omega_m=0.308$, $\Omega_\gamma=9.15\times 10^{-5}$, $\Omega_\Lambda=1-\Omega_m - \Omega_r$, and $H_0=67.8\,{\rm km}\,{\rm s}^{-1}\Mpc^{-1}$~\cite{Ade:2015xua}. $\td E_{\rm GW}$ is the total GW energy in the frequency range $(\nu,\nu+{\rm d}\nu)$ emitted in the coalescence of BHs, which is well approximated by~\cite{Cutler:1993vq,Chernoff:1993th,Zhu:2011bd}
\be
\td E_{\rm GW}(\nu) = (\pi G)^{2/3} M^{5/3} \eta \,\td \nu \times
\begin{cases}
\nu^{-1/3} &,\,\nu<\nu_1 \,,\\
\frac{\nu}{\nu_1} \nu^{-1/3} &,\,\nu_1\leq \nu<\nu_2 \,,\\
\frac{\nu}{\nu_1 \nu_2^{4/3}} \frac{\nu_4^4}{\left(4(\nu - \nu_2)^2 + \nu_4^2\right)^2} &,\,\nu_2\leq \nu<\nu_3 \,,\\
\end{cases}
\ee
where $\nu_j = (a_j\eta^2+b_j\eta+c_j)/(\pi G M)$. The coefficients $a_j,b_j,c_j$ can be found in Table I of Ref.~\cite{Ajith:2007kx}. 

The GW background when $R=R_2$ is studied in Refs.~\cite{Mandic:2016lcn,Clesse:2016ajp,Cholis:2016xvo}. We consider the case $R=R_3$ since, as shown in the previous section, it dominates over $R_2$ unless the local DM density contrast becomes very large, $\delta_0\gsim 10^{11}$. The stochastic GW background from $R_3$ has been studied in the case of monochromatic mass function in Ref.~\cite{Wang:2016ana}. 

In Fig.~\ref{GWspectrum} the predicted stochastic GW background is shown for the monochromatic mass function in blue, and for the lognormal mass function with $\sigma  = 1$ in red. Here, and for the rest of this subsection, we take $\delta_{\rm dc}=1$. For a given mass function the PBH DM fraction is fixed such that the merger rate~\eqref{rateR3} today for binaries with $m_1,m_2\in(m_{\rm min},m_{\rm max})$ is in the range inferred from the LIGO observations~\cite{Abbott:2017vtc}. The wider the PBH mass function is, the wider is the peak and the lower is the peak amplitude of the GW background. 

\begin{figure}
\begin{center}
\includegraphics[width=\textwidth]{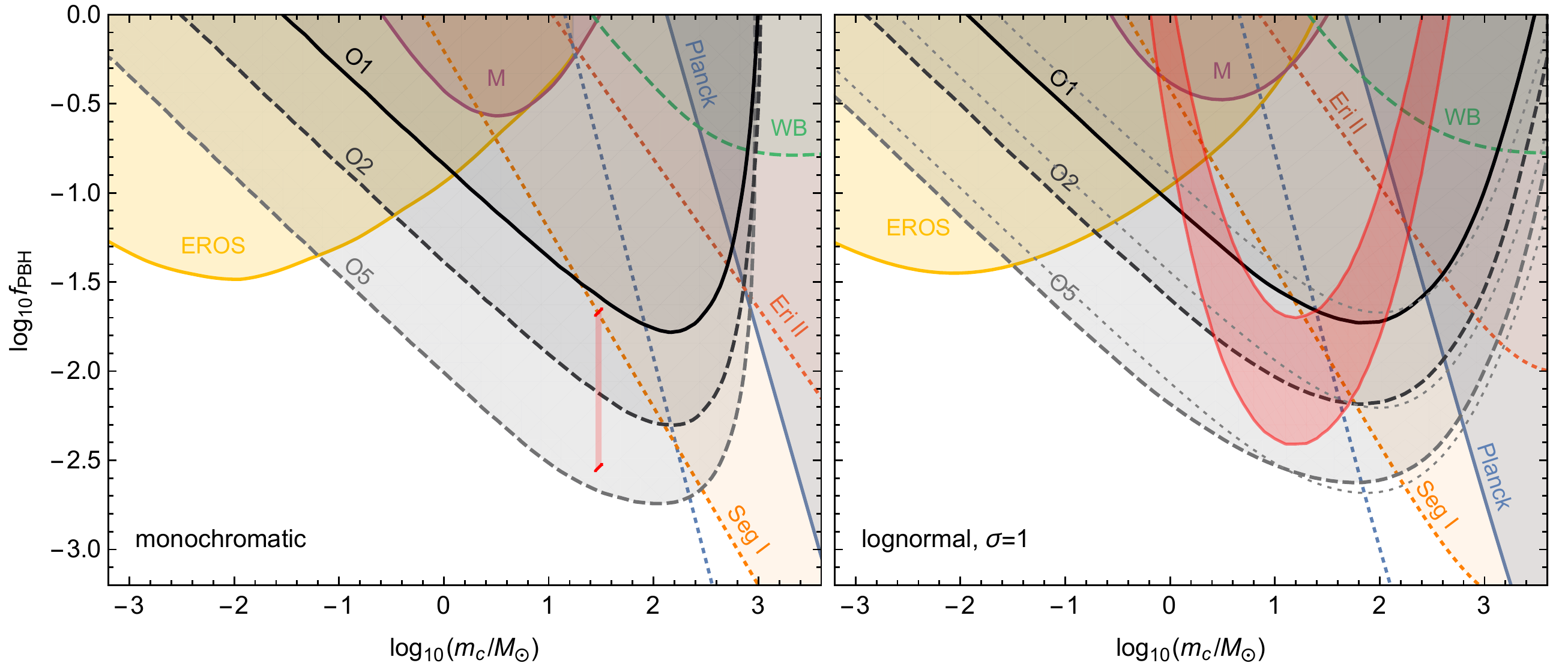} 
\end{center}
\caption{The constraints on the fraction of DM in PBHs, $f_{\rm PBH}$, from non-observation of the stochastic GW background for the monochromatic (left panel) and lognormal (right panel) PBH mass functions. The black solid line (O1) shows the constraint from the first LIGO observing run and the grey dashed lines (O2, O5) present the projected sensitivities of next phases of LIGO. The yellow and purple regions are excluded by the microlensing results from EROS~\cite{Tisserand:2006zx} and MACHO (M)~\cite{Allsman:2000kg}, respectively. The dark blue, orange, red and green regions on the right are excluded by Planck data~\cite{Ali-Haimoud:2016mbv}, survival of stars in Segue I (Seg I)~\cite{Koushiappas:2017chw} and Eridanus II (Eri II)~\cite{Brandt:2016aco}, and the distribution of wide binaries (WB)~\cite{Monroy-Rodriguez:2014ula}, respectively. On the right panel the thin dotted lines show, for comparison, the constraints calculated for the lognormal mass function from the ones in the monochromatic case by the method of Ref.~\cite{Carr:2017jsz} which has been used for all other constraints. The red lines show the values of $f_{\rm PBH}$ for which the merger rate in the LIGO sensitivity range today is $12\, \Gpc^{-3} \yr^{-1}$ (lower) and $213\, \Gpc^{-3} \yr^{-1}$ (upper).}
\label{constr}
\end{figure}

The dashed lines in Fig.~\ref{GWspectrum} show the sensitivities of LISA and LIGO GW  interferometers. The projected constraints on the fraction of DM in PBHs from non-detection of the GW background  can be adapted from these sensitivity curves. These are shown for LIGO in Fig.~\ref{constr} in the $( m_c, \,f_{\rm PBH})$ plane. Our analysis shows that the first LIGO observing run O1 has already excluded the scenario in which all the DM is in PBHs with the monochromatic mass function in the mass range from $0.03\Msun$ to $1000\Msun$. It also provides the strongest constraint on the PBH abundance so far in the mass range $0.5-16\Msun$. The expected sensitivities of the LIGO future runs, denoted by O2, O5~\cite{TheLIGOScientific:2016wyq} in Fig.~\ref{constr}, will further improve the constraint.

In the left panel of Fig.~\ref{constr} the constraints are shown for the monochromatic mass function, and in the right panel for the lognormal mass function with $\sigma=1$.  The constraint in the monochromatic case was calculated earlier in Ref.~\cite{Wang:2016ana}. Their result is of the same order of magnitude as ours, but has a different mass dependence. It is expected that the constraint becomes continuously weaker for masses far from the sensitivity range of LIGO, as is shown in our Fig.~\ref{constr}. The constraint curves in Fig. 3 of Ref.~\cite{Wang:2016ana} do not show this behaviour.

For most of the PBH observables, constraints for an extended mass function can be obtained if the corresponding constraint in the monochromatic case is known~\cite{Carr:2017jsz}. However, for PBH binary merger constraints this is not the case, because the observable depends non-linearly on the PBH mass function. In this case the constraints must be calculated separately for each mass function by comparing the GW spectrum to the sensitivity of the GW interferometer. To illustrate the resulting error, the thin dotted lines on the right panel of Fig.~\ref{constr} show the constraints obtained by the method of Ref.~\cite{Carr:2017jsz}. 

The red regions in Fig.~\ref{constr} depict the range in $f_{\rm PBH}$ as a function of $m_c$ where the present PBH merger rate for binaries with $m_1,m_2\in(m_{\rm min},m_{\rm max})$ is within the range inferred from the LIGO observations. The upper red line, which corresponds to $213\, \Gpc^{-3} \yr^{-1}$, gives an upper bound on $f_{\rm PBH}$ since a larger PBH fraction would imply a merger rate that is too large. Our analysis shows that the GW background from the PBH mergers can be probed by LIGO provided that the observed events correspond to PBH binaries. Thus the future runs of LIGO can rule out the primordial origin of the observed GW events. However, note that, as in Sec.~\ref{sec:LIGO}, we used a simplified approximation for the LIGO sensitivity. We expect that the error in the red curves in Fig.~\ref{constr} is small, and leave a more careful analysis for future work.

%-------------------------------------------------------------------------------
\subsection{Constraints from cosmology}
\label{sec4}

The CMB constraints on the stochastic GW background~\cite{Smith:2006nka,Henrot-Versille:2014jua} do not apply in case the latter is created by the PBH mergers after recombination. Nevertheless, a similar bound can be obtained by considering the cosmology of the late Universe. During its lifetime a BH binary emits maximally of order 5\% of its total mass into GWs. The BH mergers convert matter into radiation, thus a high enough merger rate could leave an imprint on the cosmological evolution of the Universe. The cosmic expansion and growth of perturbations put an upper bound $\Omega_{\rm ARB} < 0.008$ at $2\sigma$ confidence level on the total astrophysical radiation background~\cite{Torres:2016fmj}. The scenarios of decaying DM imply that no more than $3.8\%$ of DM may have decayed into dark radiation~\cite{Poulin:2016nat}.

The total change of the DM abundance due to PBH mergers after recombination at $z_{\rm CMB}=1090$ is
\be\label{eq:dDM}
	\Delta\Omega_{\rm DM} =  \int_0^\infty \frac{\td \nu}{\nu} \int_0^{z_{\rm CMB}}\td z (1+z) \frac{\td \Omega_{\rm GW}}{\td z} \,.
\ee
The fraction of DM abundance converted into radiation, 
\be
	F \equiv \frac{\Delta\Omega_{\rm DM}}{\Omega_{\rm DM}(z_{\rm CMB})},
\ee
in the late Universe is constrained by $F \lesssim 4\%$. By \eqref{eq:dDM} this implies a constraint on the stochastic GW background created after recombination. On the other hand, the $3\sigma$ tension between the low and high redshifts measurements of $\sigma_8$ and $H$~\cite{Riess:2016jrr,Hildebrandt:2016iqg,Joudaki:2016kym,Bernal:2016gxb} can be alleviated by $F\approx 2-5\%$~\cite{Chudaykin:2016yfk}, which may result form a period of intensive BH mergers.

\begin{figure}
\begin{center}
\includegraphics[height=.37\textwidth]{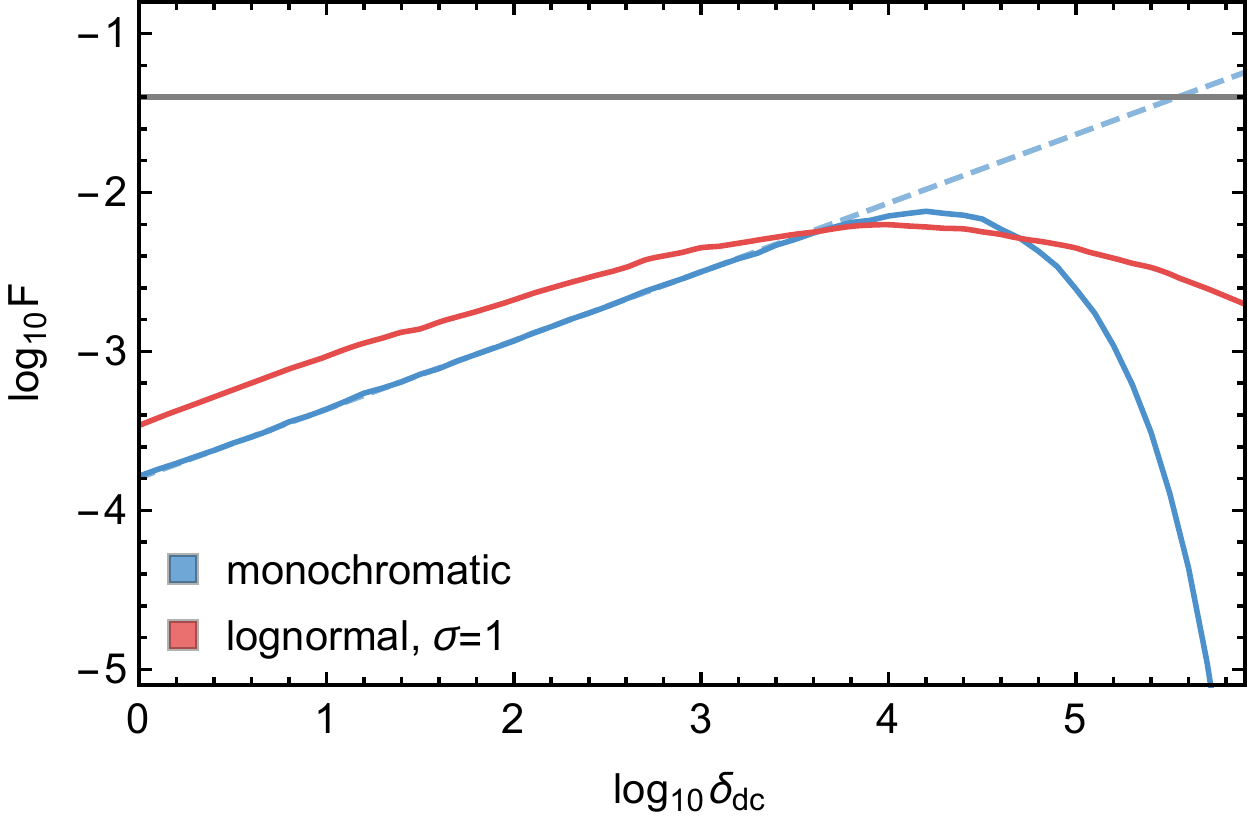}
\end{center}
\caption{The red and blue lines show the fraction $F$ of DM converted into GWs as a function of the early density contrast, $\delta_{\rm dc}$, for $m_c=30\Msun$ and $f_{\rm PBH}=1$. The dashed line shows the approximation~\eqref{zetaapprox}, and the grey horizontal line corresponds to the upper bound $F<4\%$.}
\label{Fplot}
\end{figure}

Consider first a monochromatic PBH mass function. Given that the dominant merger rate is $R_3$, the fraction of DM abundance converted into GWs can be approximated by
\be \label{zetaapprox}
	F\approx 6.3\times10^{-4} \delta_{\rm dc}^{16/37} f_{\rm PBH}^{53/37}\left(\frac{m_c}{30\Msun}\right)^{5/37} \,.
\ee
So the predicted fraction of DM abundance converted into GWs is negligible compared to the allowed fraction, i.e. $F \ll 4\%$, given that $\delta_{\rm dc}\sim1$. The fraction $F$ for $\delta_{\rm dc} > 1$ is depicted in Fig.~\ref{Fplot}. It can be seen that the approximation~\eqref{zetaapprox}, shown by the dashed line, breaks down for very large $\delta_{\rm dc}$. In that case, $F$ decreases because most of the binaries merge before recombination. The GWs originating from these mergers could still have an effect on the CMB~\cite{Smith:2006nka,Henrot-Versille:2014jua}, which we will, however, not consider in this paper. The maximal value of $F$ generally decreases when an extended mass function is considered, as illustrated in Fig.~\ref{Fplot}, and it is practically independent of the mass $m_c$. In all, $F \lesssim 1\%$ seems to provide a robust theoretical upper bound on the fraction of PBH DM that can be converted into GWs via the mechanisms considered in this work. It follows that the PBH mergers are unable to alleviate the tension of the $\sigma_8$ and $H$ measurements.

If we express $f_{\rm PBH}$ in terms of the present rate $R_3(t_0)$, \eqref{R3approx}, Eq.~\eqref{zetaapprox} can be recast as 
\be
	F \approx 1.2\times 10^{-8} \left(\frac{m_c}{30\Msun}\right) \left(\frac{R_3(t_0)}{\Gpc^{-3} \yr^{-1}} \right)\,.
\ee
Thus the merger rate inferred from the LIGO results corresponds to $F \approx 1.5\times10^{-7} - 2.6\times10^{-6} (m_c/30\Msun)$, and for the PBH mergers with masses in the LIGO sensitivity range a fraction $F \gsim 10^{-5}$ is already ruled out. For the extended mass functions these results are expected to be of the same order of magnitude.  We conclude that it seems unlikely that the PBH mergers could have presently observable cosmological consequences.

%-------------------------------------------------------------------------------
\section{Conclusions} 
\label{sec:conclusions}
%-------------------------------------------------------------------------------

We considered two formation mechanisms of PBH binaries while accounting for an extended mass function and for spatial clustering of the PBHs. The first mechanism describes binary formation in the early Universe where the PBHs have negligible peculiar velocities. The second, binary formation through GW emission, is relevant in the late Universe. We estimated the PBH merger rate and showed that the binaries created by the first mechanism dominate except in regions with an extremely high PBH density contrast. 

We assumed an extended mass function with a lognormal form. For $m_{c} = 30\Msun$ and $\sigma = 1,$ the merger rate inferred by LIGO can be reproduced if the PBH fraction is in the range $f_{\rm PBH} = 0.0045 - 0.024$, which is consistent with the present constraints on the PBH abundance. From this we conclude that the observed rate excludes the scenario in which the dominant fraction of DM is in PBHs with masses $\mathcal{O}(10\Msun)$.

We evaluated the stochastic GW background resulting from the PBH mergers and found that the merger rate from LIGO predicts a GW background that is well within the projected sensitivities of LISA and, for the PBH masses lower than $100 \Msun$, it may be detectable also by the future runs of LIGO. The non-observation of the GW background implies a constraint on the PBH abundance. Using the results from the first phase of LIGO, this yields the dominant constraint for PBH abundance in the mass range $0.5-16\Msun$. If the GW events observed by LIGO are of primordial origin, then the GW background inferred from the rate and mass function from LIGO can be detected by the future LIGO runs. A non-detection of this GW background in the future can rule out the PBH explanation of the LIGO GW events.

As the PBH mergers convert matter into radiation, they may have important cosmological implications. We find that the fraction $F$ of DM converted into GWs after recombination cannot exceed $1\%$. The bound on that quantity coming form the CMB, $F\lsim 4\%,$  is thus always satisfied, and the PBH mergers cannot alleviate the tension between the low and high redshift measurements of the Hubble constant. For the PBH mergers with masses and rates in the range inferred from LIGO we obtain $F \lsim 10^{-5}$, that is negligible.

%-------------------------------------------------------------------------------
\acknowledgments
%-------------------------------------------------------------------------------
The authors thank  B. Carr, J. Garcia-Bellido, G. H\"utsi, J. Lesgourgues, L. Marzola, V. Poulin, Y. Tada, T. Tenkanen, I. Tkachev and F. Urban for various discussions and communications on the topic. This work was supported by the grants IUT23-6 and by EU through the ERDF CoE program grant TK133.

\bibliography{citations}

\end{document}